\documentclass[%
 reprint,
 amsmath,amssymb,
 aps,
]{revtex4-2}

\usepackage{graphicx}
\usepackage{dcolumn}
\usepackage{bm}
\usepackage{braket}

\begin{document}

\title[Electric field measurement and application based on Rydberg atoms]{Electric field measurement and application based on Rydberg atoms}

\author{Bang Liu}
\author{Li-Hua Zhang}
\author{Zong-Kai Liu}
\author{Zi-An Deng}
\author{Dong-Sheng Ding}
 \email{dds@ustc.edu.cn}
\author{Bao-Sen Shi}
\author{Guang-Can Guo}
\affiliation{Key Laboratory of Quantum Information, University of Science and Technology of China, Hefei, Anhui 230026, China.}
\affiliation{Synergetic Innovation Center of Quantum Information and Quantum Physics, University of Science and Technology of China, Hefei, Anhui 230026, China.}

\begin{abstract}
Microwave sensing has important applications in areas such as data communication and remote sensing, so it has received much attention from international academia, industry, and governments. Atomic wireless sensing uses the strong response of the large electric dipole moment of a Rydberg atom to an external field to achieve precise measurement of a radio frequency (RF) electric field. This has advantages over traditional wireless sensing. The advantage of a Rydberg atom is its ultra-wide energy level transitions, which make it responsive to RF electric fields over a wide bandwidth. Here, we briefly review the progress of electric field measurement based on Rydberg atoms. The main contents include the properties of Rydberg atoms, measurement using Rydberg atoms, and experimental progress in electric field measurement in different bands. We show the different methods for detecting electric fields such as atomic superheterodyne, machine learning, and critically enhanced measurement. The development of Rydberg atomic measurement focuses on the advantages of Rydberg atomic sensing, especially compared with conventional microwave receivers. This is of major significance to developing Rydberg-based measurement in astronomy, remote sensing, and other fields. 
\begin{description}
\item[Keywords]
Rydberg atoms, microwave sensing, atomic transition, machine learning.
\end{description}
\end{abstract}

\maketitle

\section{Introduction}
\label{sec1}

Quantum mechanics was established in the early 20th century and became one of the two cornerstones of modern physics. Through quantum technologies, people have greatly improved the ability to manipulate microscopic particles and apply them to real life. Current applications of quantum technologies mainly focus on quantum communication, quantum precision measurement, and quantum computing. Quantum precision measurement has been realized in practical applications such as atomic clocks and magnetometers \cite{68}. This mainly relies on the accuracy and stability of quantum systems such as atoms and molecules, and uses them as standards to measure fundamental physical quantities with high precision. Great progress has been made in atomic-based precision measurements of atomic transition frequencies, which are now used for length and time standards. Atomic clocks have now achieved higher precision than $10^{-19}$ \cite{1,74,75,76}. Progress has also been made in measuring weak magnetic fields up to a precision of fT/Hz$^{1/2}$ \cite{2,3,4,5}, which promotes applications in encephalopathy and cardiac magnetism measurement. Recently, there has been increased interest in measuring microwave electric fields \cite{6,7,8} using Rydberg atoms.

A Rydberg atom is a special atom with at least one electron in a highly excited state, which generally requires a principal quantum number $n > 20$. The large electric dipole moment of a Rydberg atom makes it couple strongly to a weak electric field, which can be used to sense electric fields. When an electric field is applied, the Rydberg atom couples to this external field through its large electric dipole moment, and the state of the Rydberg atom changes. The atomic energy levels move and the population of each energy level changes. This change in the atomic state is monitored using electromagnetically induced transparency (EIT) \cite{9}, which is an all-optical readout of the Rydberg atomic state \cite{60,77}.

For EIT, the probe light can completely pass through the atomic medium within a three-energy-level atomic system interacting with the optical field, which can be explained using quantum optics. Generally, a light beam passing through an atom will be absorbed if its frequency resonates with an atomic energy level, whereas the EIT appears in a three-energy-level system in which the presence of another coupling light causes the probe light to pass completely through the atomic medium as if the atom becomes transparent. In an EIT, a dark atomic state is formed. This special state does not couple to the probe light, which causes the EIT. However, an applied electric field affects the dark state and thus the EIT spectrum of the atom. Thus, for a Rydberg EIT \cite{71,73}, we can use this technique to detect the change of the Rydberg state when the electric field couples with the atom. Then we obtain the electric field information  through detecting the changes of the measured EIT spectrum, including energy level shifts or population changes of the Rydberg atom. Because of the precision of the spectral measurement, the electric field can be measured with high precision.

Compared with traditional antenna measurement, electric field measurement based on Rydberg atoms has the following advantages because it uses the quantum properties of atoms:

\begin{itemize}

\item High sensitivity: Because the large electric dipole moment of a Rydberg atom is extremely sensitive to the external electric field, electric field measurement using the Rydberg atom is actually a quantum measurement scheme, and its quantum projection limit far exceeds that of traditional measurement methods.

\item Low noise and immunity to electromagnetic interference: Traditional electric field measurements are based on antennas, which are inevitably affected by thermal noise. For measurement using Rydberg atoms, the atoms packed in glass vapor cells have no electronic components and are immune to thermal noise.

\item Large bandwidth: Because of the abundance of energy levels in Rydberg atoms, it is easy to change these energy levels to couple with electromagnetic waves in a wide range of frequencies from DC to the THz band. This enables sensing and measurement of electric fields in these frequency bands without changing the experimental setup.

\item Traceability: For measurement using Rydberg atoms, the optical response properties are traceable to a number of fundamental physical constants. This means that the electric fields measured with Rydberg atoms do not require calibration and can be measured directly.

\item Portability and integration: The core component for sensing the electric field using Rydberg atoms is a vapor glass cell, which can be on the order of cm in size and is easily carried, integrated, and miniaturized.

\end{itemize}

Electric field measurement has important applications in real life. In particular, an electric field in the microwave band has strong penetration and long propagation distance, and can carry a large amount of information. This gives microwave measurement technology a lot of potential in a wide range of applications. Thus, the international research community, industry, and governments have paid great attention to it. For example, microwave technologies play a very important role in data communication and remote sensing, such as in 5G communication and radar detection. Therefore, electric field measurements based on Rydberg atoms can not only be used to establish atom-based electric field measurement standards, but also have new practical applications.

This paper introduces the basic principles and applications of electric field measurements based on the Rydberg atom, and is organized as follows. Section \uppercase\expandafter{\romannumeral2} introduces the properties of the Rydberg atom and the basic concepts and principles of EIT and the Autler–Townes (AT) effect in electric field measurements. Section \uppercase\expandafter{\romannumeral3} presents the basic schemes and principles of using the Rydberg atom to measure electric fields in the frequency range from DC to the THz region. Section \uppercase\expandafter{\romannumeral4} focuses on some applications based on Rydberg atomic measurements of electric fields, especially sensing and communication. Finally, an outlook is presented on Rydberg atomic measurements of electric fields and their applications.

\section{Rydberg atomic properties}
\label{sec2}

\subsection{Rydberg Atom}

The Rydberg state is a highly excited electronic state formed by the transition of the outermost electrons of an atom or molecule to an orbital with a higher principal quantum number $n$ \cite{10}; an atom in the Rydberg state is called a Rydberg atom. Because the outermost electron in the Rydberg atom is far from the nucleus, the atom can be regarded as hydrogen-like. 

The physical properties of Rydberg atoms can be expressed using simple formulas similar to those of hydrogen atoms. However, Rydberg atoms have the following special properties: 

1. Large atomic radius. According to Bohr's model of the hydrogen atom, the orbital radius of the electron is $r\propto n^2$, where $n$ is the principal quantum number. The principal quantum number of the outermost electron of the Rydberg atom is usually large, so the radius of the Rydberg atom is large. 

2. Rydberg atoms are easy to ionize. The ionization energy of an alkali metal atom is expressed as $E_{nLJ}=-hcR^{*}\text{/(\ensuremath{n^{*})^{2}}}=-hcR^{*}/(n-\delta_{nLJ})^{2}$, where $R^*$ is the modified Rydberg constant and $n^*$ is the effective principal quantum number, which can be expressed as $n^*=n-\delta_{nLJ}$ ($\delta_{nLJ}$ represents the quantum defect for the different quantum numbers $n$, $L$, and $J$). The ionization energy of a Rydberg atom is low for large principal quantum number $n$. 

3. The energy spacing of Rydberg atoms is small. The energy difference between two energy levels is $\Delta E\propto1/n^{2}-1/(n+1)^{2}\approx2/n^{3}$, where the principal quantum numbers are $n$ and $n + 1$, respectively. The energy level difference between two adjacent Rydberg states is small because $n$ is generally large. 

4. Long lifetime. The lifetime of atoms in lower excited states is generally $10^{-8}$ s, and the lifetime of highly excited Rydberg atoms can reach the order of microseconds or even milliseconds.

\begin{table}[htbp]
    \centering
    \caption{Properties of Rydberg atoms}
    \label{tab1}
    \begin{tabular}{|c|c|c|}
        \hline
        Property & $n$ dependence & Na (10d) \\
        \hline
        Binding energy & $n^{-2}$ & 0.14 eV \\
        \hline
        Energy spacings & $n^{-3}$ & 0.023 eV \\
        \hline
        Orbital radius & $n^{2}$ & 147 $a_{0}$ \\
        \hline
        Geometric cross section & $n^{4}$ & 68000 $a_{0}^{2}$ \\
        \hline
        Dipole moment ($\bra{nd}er\ket{nf}$) & $n^{2}$ & 143 $ea_{0}$ \\
        \hline
        Polarizability & $n^{7}$ & 0.21 MHz $cm^{2}/V^{2}$ \\
        \hline
        Radiative lifetime & $n^{3}$ & 1.0 $\mu$s \\
        \hline
        Fine-structure interval & $n^{-3}$ & -92 MHz \\
        \hline
    \end{tabular}
\end{table}

As mentioned above, the Rydberg atom has a large radius, high energy, small energy level difference, and long lifetime. The electric dipole moment of the atom is also large owing to its relatively large radius, which makes the Rydberg atom very sensitive to external electric fields. Meanwhile, the small neighboring energy level difference in the Rydberg atom just corresponds to the microwave band. Because of the richness of Rydberg atomic energy levels, changing the principal quantum number $n$ covers a very wide frequency range from low frequency to the THz order, which confers great advantages in electric field measurements.

\subsection{Detection Mechanism}

EIT is a coherent optical nonlinear phenomenon first discovered experimentally by Harris and his co-workers \cite{11}. It makes the medium transparent in a narrow spectral range around the absorption line. Rydberg-EIT is a real-time non-destructive method that can directly detect the population of Rydberg atoms. In this section, we introduce the mechanism of measurement through the Rydberg-EIT method \cite{7,8}. 

\begin{figure*}[htbp]
\centering
    \includegraphics[width=0.9\linewidth]{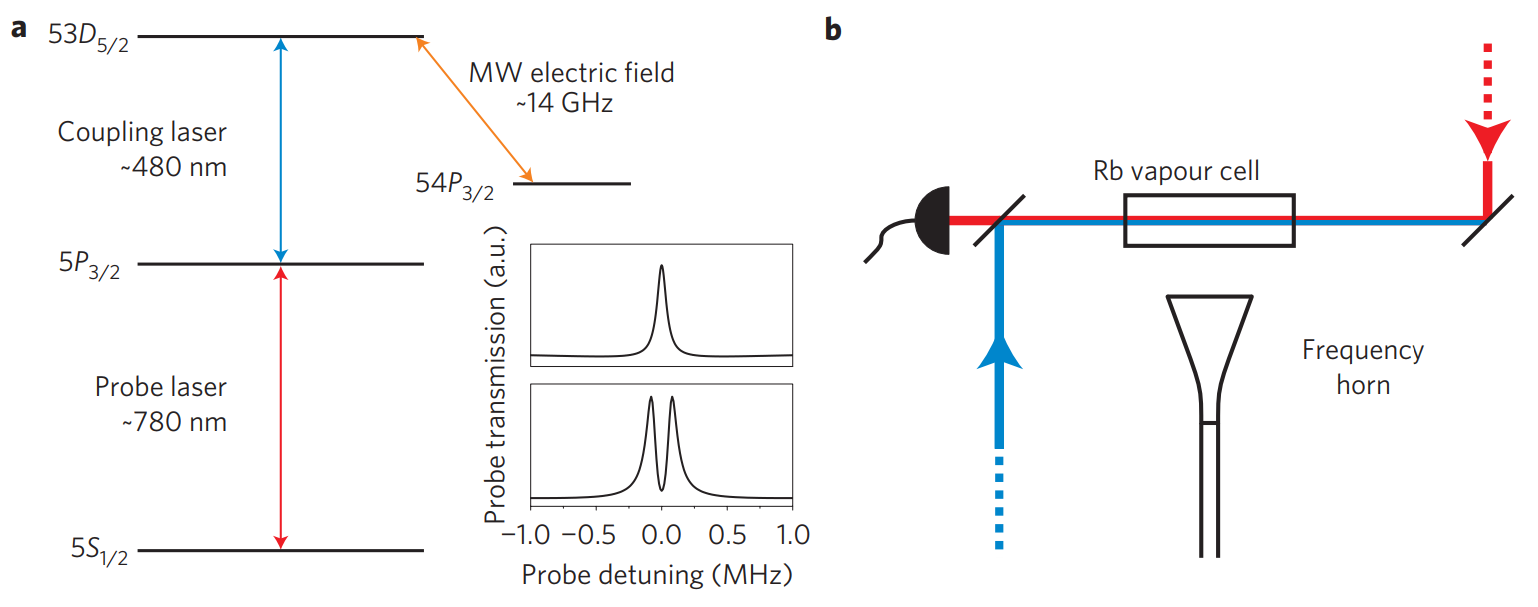}
\caption{(a) Diagram of a four-level system used to sense the microwave electric field. The top of the illustration shows an example of an electromagnetically induced transparency (EIT) associated with a three-level system without a microwave electric field. The bottom of the illustration shows an example of a bright resonance generated within the EIT window when a microwave electric field is present. (b) Experimental apparatus \cite{8}.}
\label{fig1}
\end{figure*}

The emergence of Rydberg-EIT is actually from the dark states that make it impossible for atoms to absorb the probe light. Examining the Hamiltonian eigenstates of the three-level EIT system, we find that the energy of an eigenstate is 0. This eigenstate wave function can be expressed as
\begin{equation}
    \ket{D} \propto -\ket{1}+\ket{3}, \label{1}
\end{equation}
where $\ket{1}$ and $\ket{3}$ are the ground state and Rydberg state, respectively. The eigenstate $\ket{D}$ is called the dark state because its intermediate excited state has zero population. That the excited state has zero population means no atoms absorb the probe photons through the excited state, i.e., the probe light then passes directly through the atomic vapor without absorption. The dark state is the direct reason for the Rydberg-EIT effect. Thus, we see that there is a transmission window near the central frequency of the probe light, as shown in Fig.~\ref{fig1}(a), and this corresponds to the dark state.

A microwave field can be added to couple the transition between the uppermost energy level and the fourth energy level, as shown in Fig.~\ref{fig1}(a). This splits the transmission spectrum of the probe light, and two transmission peaks appear. When the Rabi frequency of the signal microwave field is much higher than the Rabi frequencies of the probe light and the coupled light, the Rabi frequencies of the probe light and coupled light can be neglected. A pair of microwave-decorated dark states is generated at this time with an eigenenergy difference of
\begin{equation}
    \Delta E = \hbar \Omega, \label{2}
\end{equation}
where $\Omega = \sqrt{\Omega_s^{2}+\Delta^{2}}$ is the effective Rabi frequency, $\Delta$ is the detuning of microwave, $\Omega_s$ is the Rabi frequency of the signal microwave field, and $\hbar$ is Planck's constant. The transmission spectral line is split by the Autler–Townes (AT) effect. The splitting interval $\Delta f$ of the microwave-decorated dark state can be obtained from spectral measurements, and therefore the electric field strength of the signal microwave field can be inferred from it. In the case of detuning, the relationship between the splitting interval and the Rabi frequency of the microwave field is
\begin{equation}
   \Delta f=k \Omega=k \sqrt{\Omega_s^2+\Delta^2}, \label{3}
\end{equation}
where $k$ is the correction factor introduced to take into account the Doppler effect in the hot atomic system, and is 1 for scanning coupling light and $\lambda_{p}\text{/}\lambda_{c}$ for scanning probe light. We obtain
\begin{equation}
   E=\frac{ \hbar \Delta f}{ \mu_r}, \label{4}
\end{equation}
where $\mu_r$ is the corresponding dipole moment. This is the basic principle of measuring the microwave electric field using the AT splitting of the Rydberg-EIT spectrum.

\subsection{Stark Shift}

In this section, we introduce the mechanism of interaction between atoms and an applied electric field, which is the most basic principle for sensing electric fields with atoms. In traditional measurement, an external electric field induces a current in the antenna, and then a series of circuits amplifies the signal to measure the electric field. The most significant effect of atoms in an applied electric field is the Stark Shift, which is also a quantum effect, first discovered by Stark in 1913.

Under a DC electric field, an atomic energy level undergoes an energy shift under a DC electric field. When the electric field strength is $E$, an atom with an intrinsic electric dipole moment undergoes an energy shift $\delta E=\boldsymbol{d}\cdot \boldsymbol{E}$, which is called the linear Stark effect. If the intrinsic electric dipole moment is 0, an electric dipole moment $\boldsymbol{d}=\alpha \boldsymbol{E}$ is induced under the electric field, where $\alpha$ is the polarizability of the atom. The energy level shift is thus $\delta E\propto\alpha E^2$, which is called the squared Stark effect \cite{14,15}.

For an AC electric field, the effect is different depending on the electric field frequency. A non-resonant weak electric field has an effect similar to that of direct current; that is, energy level shifts occur. A strong field induces sidebands in the spectrum that are explained by Floquet theory \cite{16,17,18}. Resonant electric fields induce AT splitting, as described above.

In short, atoms have different perturbations under an external electric field that result in different responses in the EIT spectrum. We need to record details of the EIT spectrum to get information on the electric field.

\section{Electric field measurement}
\label{sec3}

\subsection{DC Electric Field Measurement}

The polarizability of a Rydberg atom is proportional to $n^7$, and the electric field strength threshold for field ionization is proportional to $n^{-4}$ \cite{10}. Therefore, atoms in the Rydberg state are very sensitive to the electric field, and the magnitude of the weak electrostatic field can be measured by recording the Rydberg atomic level shift through microwave \cite{21,22,61} or laser \cite{23,24} spectroscopy. The Stark shift of the Rydberg state is measured from the vacuum UV-millimeter double-resonance spectrum of the high Rydberg state of the krypton atom \cite{22}. As discussed in the previous section, the Rydberg atom produces a Stark energy shift $\delta E=\alpha E^2$ in an external DC electric field, so the magnitude of the applied electrostatic field can be obtained by simply measuring the energy shift through the spectrum. The minimum electrostatic field strength that can be measured is $\pm20$ $\mathrm{\mu V/cm} $ \cite{22}.

Rydberg atoms are detected through field ionization, so this is a destructive detection scheme and does not allow for real-time continuous electric field measurement. In addition, this measurement scheme is mainly limited by the resolution of the acquired spectrum; thus, a higher-energy UV laser is required to excite the atoms to higher Rydberg states.

\subsection{Low-frequency Electric Field Measurement (kHz)}

\begin{figure}[htbp]
\centering
    \includegraphics[width=0.8\linewidth]{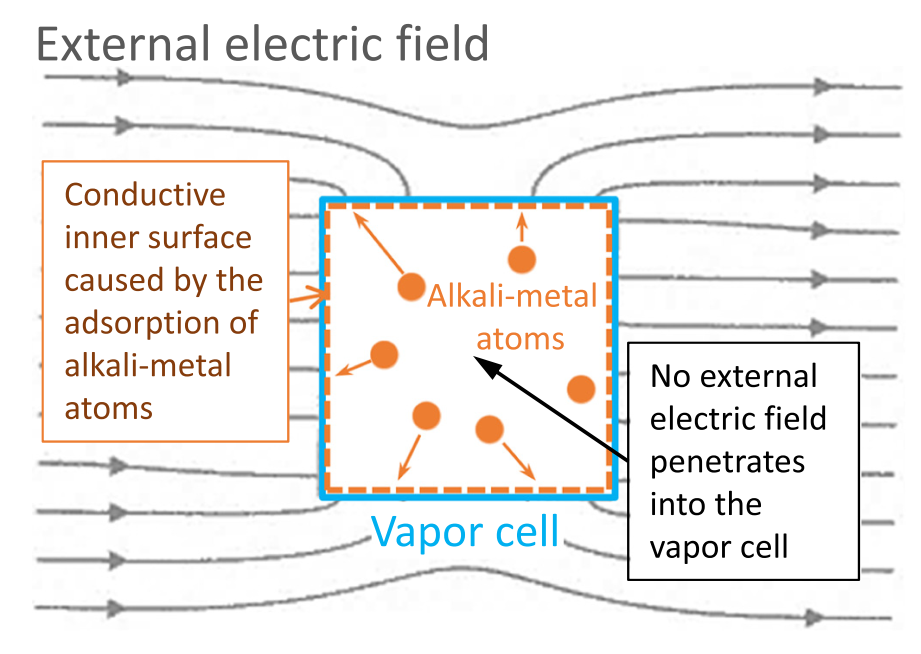}
\caption{Adsorption of alkali metal atoms causes the inner surface of the vapor cell to be slightly conductive. When a low-frequency external electric field is applied, the free charge on the inner surface is redistributed to maintain an equipotential state, and no electric field enters the vapor cell \cite{32}.}
\label{fig2.2}
\end{figure}

A low-frequency electric field on the order of kHz mainly causes a Stark shift when interacting with Rydberg atoms; thus, it is still necessary to measure the Rydberg energy level shift to obtain information on the electric field. However, the atoms are usually held in a glass container, and the atomic species are usually alkaline metals such as Rb and Cs, as shown in Fig.~\ref{fig2.2}. These alkali metal atoms will be adsorbed on the inner surface of the glass container, causing low-frequency electric field shielding \cite{25,26}. Only when the electric field oscillates at high frequencies (from the sub-MHz to the THz orders) can it penetrate the glass cell and be detected by the atoms \cite{27,28,29,30,31}. Therefore, the electromagnetic shielding effect of the glass vapor needs to be considered when measuring low-frequency electric fields. 

For this purpose, researchers have used a Rb gas vapor cell made from sapphire single crystals with very slow electric field shielding times, up to the order of seconds \cite{32}. Low-frequency electric fields have been measured using a sapphire gas vapor cell with a light-sensing internal bias electric field.
This is sensitive enough to respond linearly to an external electric field. Such a sensor can be used to monitor weak ambient AC electric fields and other electronic noise, and to remotely detect moving charged objects. The main sensitive unit of this atomic sensor uses an effective volume of 11 $\rm mm^3$ and provides a scattering noise limit of approximately 0.34 mV/m/Hz$^{1/2}$ with a low 3-dB cutoff frequency of approximately 770 Hz \cite{32}.

This approach enables additional applications in atomic electric field sensing such as calibration of DC and low-frequency electric fields, communication in the extremely-low-frequency (ELF) and ultra-low-frequency (SLF) bands (kHz or below), activity detection through remote sensing, geoscience, and bioscience.

\subsection{Intermediate-frequency Electric Field Measurement (MHz)}
\subsubsection{Intermediate-frequency Strong-field Measurement}
\begin{figure}[htbp]
\centering
    \includegraphics[width=0.8\linewidth]{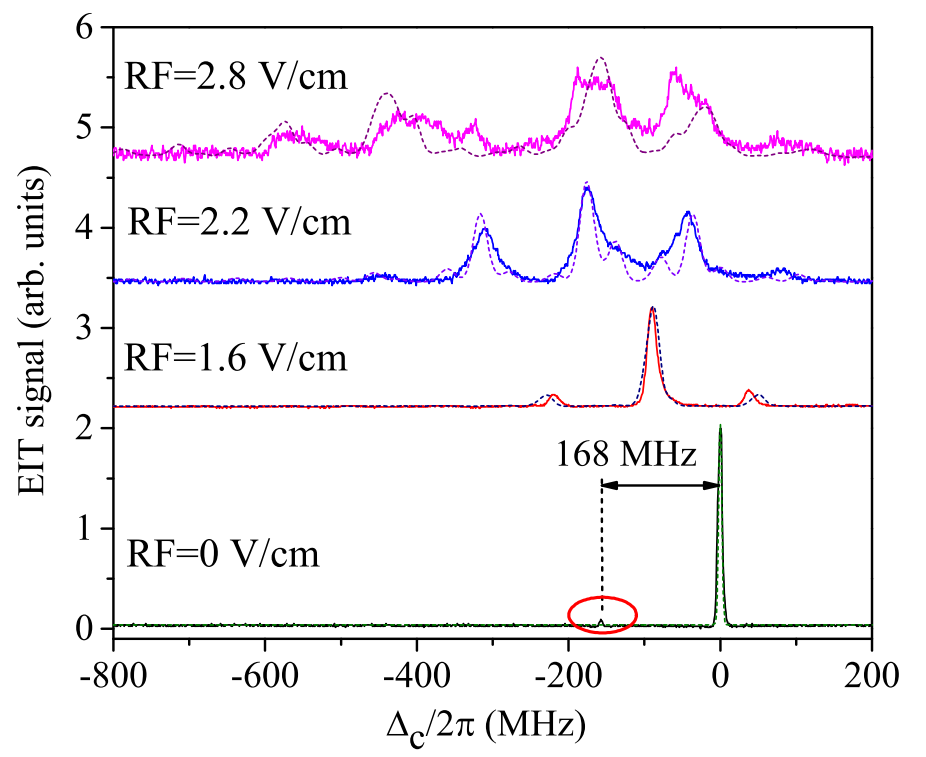}
\caption{Measured (solid line) and calculated (dashed line) Rydberg EIT spectra under RF modulation (modulation frequency $\omega_m$ = 2$\pi$ × 70 MHz) and the indicated RF field amplitudes \cite{33}.}
\label{fig2.3}
\end{figure}

For an intermediate-frequency (IF) electric field in the MHz band, there is almost no resonance with the Rydberg atomic energy levels; thus, the coupling with the Rydberg atoms is mainly non-resonant. A weak applied MHz field leads to a Stark shift, and the energy level shift is proportional to the square of the electric field amplitude. However, for a strong MHz field, a general perturbation calculation fails owing to the relatively strong interaction \cite{55,56,57,62}. In this case, the Floquet method is used to solve for the response of the atoms to the electric field. The Rydberg level in this case is modulated by the applied electric field, and a series of discrete energy levels is produced in which the spacing between levels equals the frequency of the applied electric field (in natural units). The energy levels of an atom are
\begin{equation}
    \delta_N=E^{(0)}-\frac{\alpha^2}{2}E^2_{dc}-\frac{\alpha^2}{4}E^2_{ac}+N\hbar \omega_m, \label{5}
\end{equation}
where $E^{(0)}$ corresponds to the energy of the atomic state, $E_{dc,ac}$ represents the amplitudes of the DC and AC fields, $N$ is the sideband order, and $\omega_m$ is the frequency of the applied electric field. The EIT spectrum exhibits the sideband characteristics shown in Fig.~\ref{fig2.3}. The interval between the sidebands and the main peak reflects the electric field frequency, and the signal height reflects the electric field strength. By comparing the calculated spectrum with the experimental spectrum, we can obtain the electric field strength with an uncertainty of 3\%, corresponding to a minimum measurable electric field of 1 V/cm \cite{24,33,34}.

In strong MHz fields, adjacent hydrogen-like manifolds begin to intersect displacement levels, providing a rich spectral structure suitable for precision field measurements. Each energy level position $\omega_{v,N}$ and transition probability $S_{v,N}$ can be calculated as
\begin{gather}
    \hbar \omega_{v,N}=W_v+n\hbar \omega_{RF}, \label{6} \\
    S_{v,N}=(eF_L/\hbar)^2 \lvert \sum_k \tilde{C}_{v,k,N}\cdot \bra{k} \hat{r} \ket{5P_{3/2},m_J} \rvert^2. \label{7}
\end{gather}
More details can be found in \cite{34}. An RF electric field is determined by matching observed spectral signatures including AC level shifts, even harmonic RF sidebands, and RF-induced avoidance crossings in the Rydberg manifold, with spectra calculated from non-perturbative Floquet theory. The measurement precisions for RF field frequency and electric field amplitude are 1.0\% and 1.5\%, respectively, and the maximum field strength reaches 5 kV/m \cite{35,36}.

\begin{figure*}[htbp]
\centering
    \includegraphics[width=0.95\linewidth]{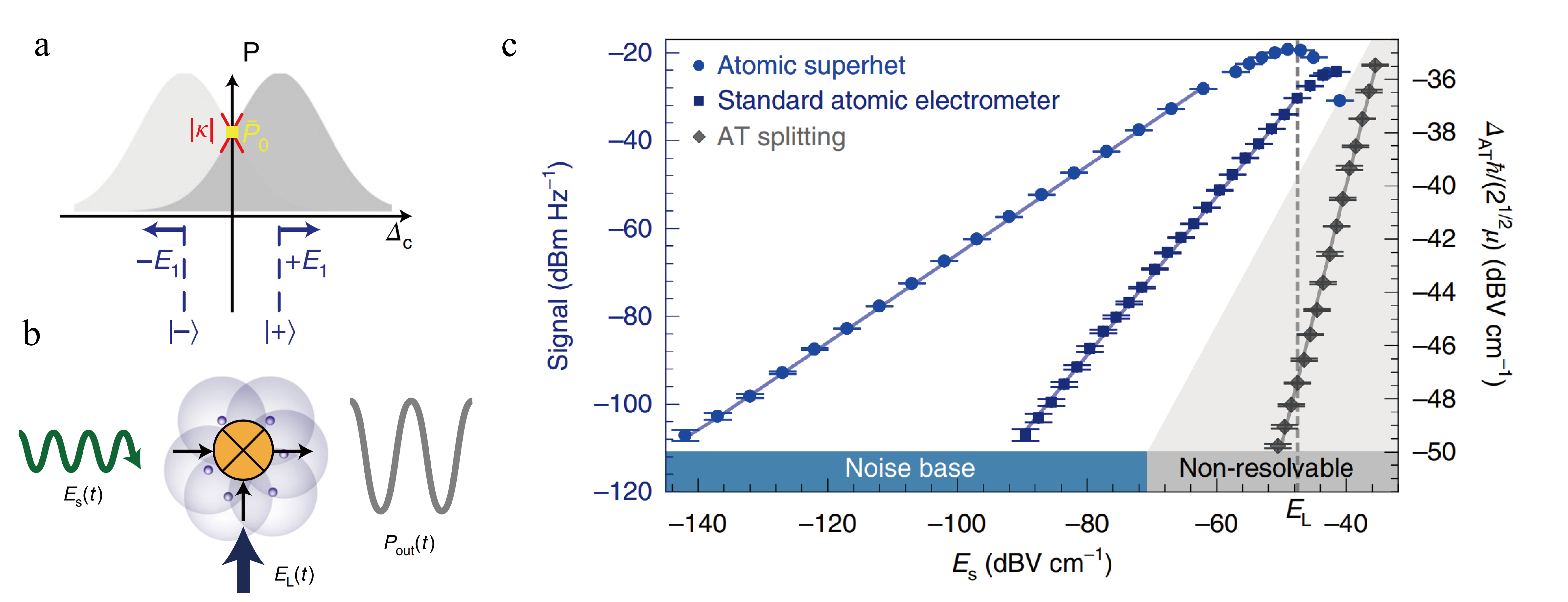}
\caption{(a) and (b) Schematics of the principles of a superheterodyne Rydberg atomic receiver, where $E_s$ is the amplitude of the signal microwave field to be measured, $E_L$ is the amplitude of the local oscillating microwave field, and the output signal $P_{out}$ is read out through EIT spectroscopy. (c) Comparison of the sensitivities of atomic electric field meters and atomic superheterodyne measurement systems \cite{7}.}
\label{fig2}
\end{figure*}

\subsubsection{Intermediate-frequency Weak-field Measurement}

The previous section described how the interaction between an IF strong field and a Rydberg atom causes modulation sidebands in the EIT spectrum and more complex energy level crossings. These sidebands and the information on the crossing can be used to determine the electric field strength. However, this method is limited to intermediate to strong fields. When a weak IF electric field is applied, the perturbation to the Rydberg atoms is very small, and only a small perturbation is generated in the spectrum. This spectral resolution limit makes it difficult to measure such small disturbances. To this end, we discuss a measurement scheme using a superheterodyne receiver, where a local oscillator field is introduced to amplify the system response to the weak IF electric field and measure the field. At an electric field frequency of 30 MHz, a minimum field strength of 30 µV/cm is obtained with a sensitivity of up to $-65$ dBm/Hz and a linear dynamic range of over 65 dB \cite{53}.

The long wavelength and long propagation distance of a MHz RF electric field make it very significant in fields such as short-wave international and regional broadcasting and aviation air-to-ground communication. The main advantage of using Rydberg atoms to measure an IF electric field is that the volume of the sensing unit can be made very small, reaching the order of cm. Meanwhile, Rydberg atom-based electric field sensors enable measurements from weak fields on the order of µV/cm to strong fields on the order of kV/m. Because of these advantages, electric field sensors based on Rydberg atoms are expected to be used in more applications such as long-distance communication, over-the-horizon radar, and RF identification (RFID).

\subsection{High Frequency Electric Field Measurement (GHz)}

For a microwave electric field with a frequency exceeding 1 GHz, resonance interaction with the Rydberg atom occurs because the frequency is close to the adjacent energy levels of the Rydberg atom, resulting in AT splitting in the EIT spectrum. This effect can be used to measure electric fields, which we will describe in detail below.

\subsubsection{Measurement of a Microwave Electric Field through AT splitting}

AT splitting of a Rydberg atom was first used to measure a microwave electric field by James Shaffer's group at the University of Oklahoma in 2012 \cite{8}. A sensitivity of 30 ${\rm \mu V/cm/Hz^{1/2}}$ was obtained and the minimum detectable field is as low as  8 ${\rm \mu V/cm}$. The electric field detection is shown in Fig.\ref{fig1}.

Subsequent improvements have been made using active control of frequency-modulated spectroscopy and residual amplitude modulation (AM) to improve the signal-to-noise ratio of the optical readout of Rydberg atom-based RF electrometry \cite{69,70,72}. A sensitivity up to 3 ${\rm \mu V/cm/Hz^{1/2}}$ can be achieved, reaching the photon shot noise limit \cite{37}.

\subsubsection{Atomic Superheterodyne Measurement }

\begin{figure*}[htbp]
\centering
\includegraphics[width=0.9\linewidth]{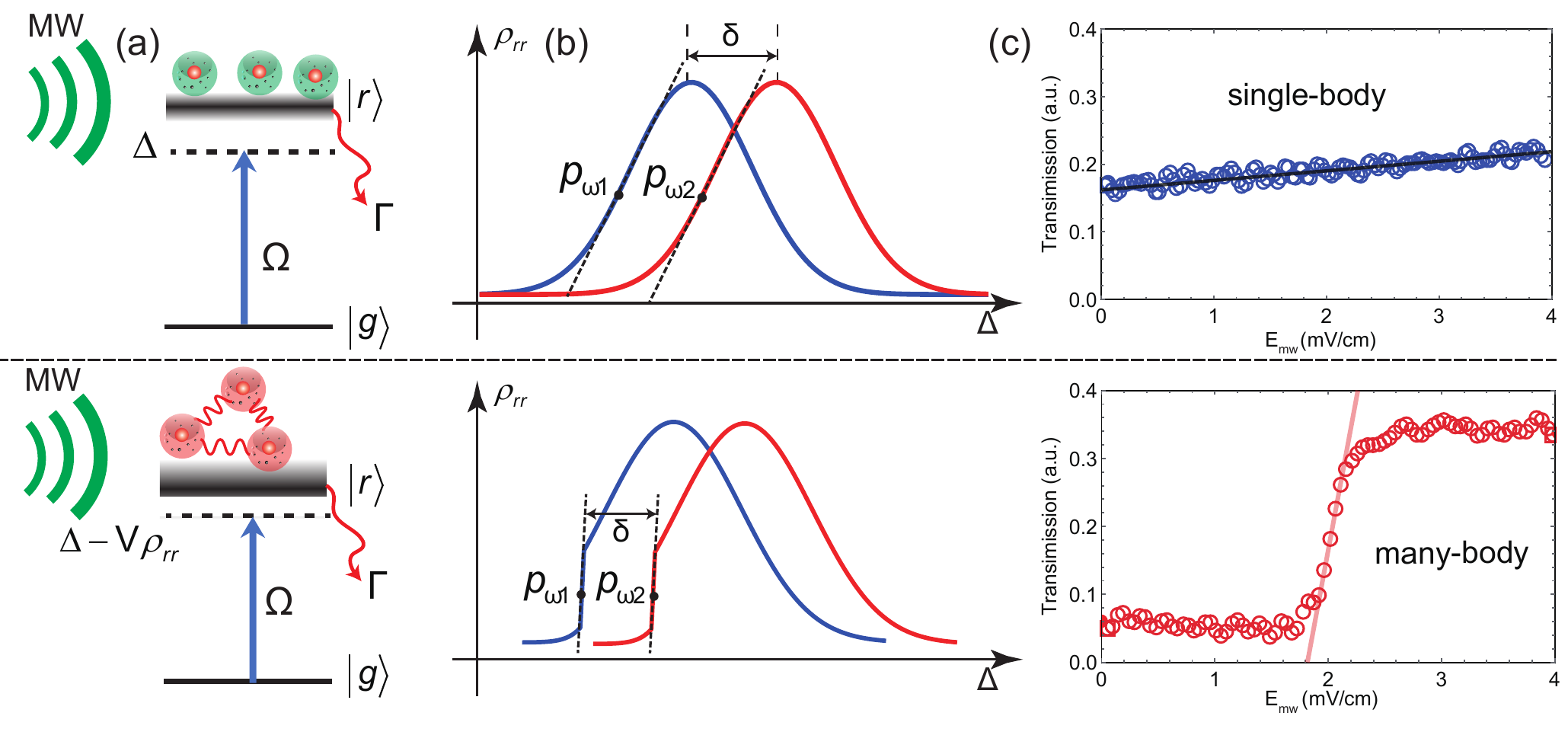}
\caption{\textbf{Principles of single-body [top row] and many-body [bottom row] Rydberg metrology.} (a) Energy diagram for a two-level atom showing the ground state $\left| g \right\rangle$ and Rydberg state $\left| r \right\rangle$. In the many-body case, the Rydberg resonance is modified by the many-body interaction strength, $V = {C_6}/{r^6} $.
(b) The blue and red curves respectively represent the spectrum with and without an external microwave field, which induces a shift $\delta$. The measurement sensitivity is highest when the derivative $\mathrm{d}\rho_{rr}/d\Delta$ is maximal, as indicated by points $p_{w1}$ and $p_{w2}$. The steeper slope near the critical point in the many-body case [bottom row] results in enhanced measurement sensitivity. (c) Transmission under different microwave field amplitudes for the single-body and many-body cases. The steep edge at $E_{mw}=$2.0 $\mathrm{ mV/cm}$ indicates that the tiny variance of the microwave field induces a giant change in the transmission, and thus the sensitivity of the many-body case is higher than that of the single-body case \cite{Ding_2022}.
}
\label{Enhanced}
\end{figure*}

Using the AT splitting of the spectrum to measure the electric field has certain limitations. This is because the derivation requires the Rabi frequency of the microwave electric field to be strong enough. That is, the electric field strength is relatively strong, so AT splitting is only suitable for measuring intermediate and strong fields. Because a weak field is not enough to split the spectrum, it only reduces the EIT peak. Therefore, new methods are needed to measure a weak field. Next, we describe how introducing a local oscillator electric field into a superheterodyne system greatly improves the sensitivity of measuring electric fields.

The Rydberg-superheterodyne process is as follows.
The local oscillator signal $E_L$ of a strong near-resonant microwave produces AT splitting in its spectrum. This signal and the signal to be measured, $E_S$, pass through the Rydberg atom. The interaction mixes frequencies, as shown in Fig.~\ref{fig2}, and the resulting IF signal $P_0$ is loaded onto the EIT spectrum.

The strong local oscillator microwave field causes the atom to produce two $\hbar\Omega_L$ dressed states with an energy interval of $\ket{\pm}$, where $\Omega_L$ is the Rabi frequency of the local oscillator field. When the very weak signal field is loaded on the atom, the energy shift $\ket{\pm}$ of the two dressed states is
\begin{equation}
    \pm E_1 = \pm \hbar \Omega_s \mathrm{cos}(2 \pi \delta_s + \phi_s)/2, \label{8}
\end{equation}
where $\Omega_s$ is the Rabi frequency of the signal field, $\delta_s$ is the amount of frequency detuning of the signal field relative to the local oscillator field, and $\phi_s$ is the phase difference between the signal and the local oscillator. The energy shift of these two dressed states results in a linear change in transmittance at the center of the EIT spectrum (at the resonance). The relationship between the transmittance change $P_{out}(t)$ and the signal electric field is
\begin{equation}
    P_{out}(t)=\lvert P(\delta_s) \rvert \mathrm{cos}(2\pi\delta_st+\phi_s).
\end{equation}
Here, $P_{out }  (t)=P(t)-\overline{P}_0$ is the change of  transmittance relative to the transmittance $\overline{P}_0$ under a no-signal electric field and $|P(\delta_s)|$ is the single-sided Fourier spectrum corresponding to a frequency of $\delta_s$. The transmittance change is proportional to the Rabi frequency of the signal field: 
\begin{equation}
    \Omega_{s}=|P(\delta_{s})|/|\kappa_{0}|.
\end{equation}

The intensity of the signal microwave electric field can be measured by measuring the intensity of the beat frequency. This scheme was successfully demonstrated experimentally by a research group at Shanxi University \cite{7} with a sensitivity reaching 55 ${\rm nV/cm/Hz^{1/2}}$. The minimum detectable field is three orders of magnitude lower than what can be achieved with AT splitting. This Rydberg atomic superheterodyne receiver allows traceable International System of Units (SI) measurements with an uncertainty of $10^{-8} {\rm V/cm}$ and also allows phase and frequency detection. This approach makes it possible to measure microwave electric fields using Rydberg atoms with much higher sensitivity, greatly enhancing the use of Rydberg atoms to measure weak electric fields.

\subsubsection{Enhanced Precision via the Critical Point of the Phase Transition in Rydberg Atoms}

Because of the large interaction volume and the strong many-body interaction, Rydberg atoms near the phase transition point produce an avalanche for a very small disturbance. That is, the atoms in the many-body Rydberg state excite surrounding atoms to many-body Rydberg states, thus amplifying the effect of small perturbations on the Rydberg atoms. This property can be used to further enhance the precision and sensitivity of Rydberg atoms to detect microwave electric fields by detecting the steep optical spectral of Rydberg atoms to microwaves near the phase transition point \cite{PhysRevA.78.042105,PhysRevA.93.022103,PhysRevLett.126.200501}.

There is a discontinuous phase transition in Rydberg atoms that is characterized by a sharp edge in the transmission spectrum \cite{PhysRevLett.111.113901,ding2020phase}. The phase transition is caused by the avalanche generated by the interaction between Rydberg atoms, which makes the number of Rydberg atoms change significantly at the critical point of the phase transition. This critical phenomenon was used to sense a microwave electric field in recent work \cite{Ding_2022}. If an external microwave electric field is applied at this time, the spectral lines move because of the Stark effect, as shown in Fig.~\ref{Enhanced} (b). The shift of the steep edge is more obvious than the overall movement; that is, the change in the external field is more obvious at the edge. Compared with using the overall spectral shift, using the critical point of the phase transition to measure the microwave electric field is equivalent to using a ruler with a finer scale, so the field can be measured with higher precision.

\subsubsection{Microwave-to-Optics Conversion through Rydberg Atoms}
Nonlinear optical processes such as four-wave and six-wave mixing can be realized using the atomic energy level structure. Microwave photons in these nonlinear optical processes can be converted to optical bands by choosing suitable energy levels \cite{48,49,50,51}. The group at South China Normal University has realized six-wave mixing using a cold atomic system to efficiently convert microwave photons to optical bands through microwave coupling of the transition of two neighboring Rydberg energy levels \cite{52}. The experiment is based on a two-dimensional cold atomic system with large optical depth, and the conversion efficiency reaches 82\% with a bandwidth close to 1 MHz. This reliably converts single-photon quantum states from the microwave domain to the optical domain. This mechanism can be used to convert dozens of microwave photons to the optical band. A practical example is using single-photon counters to measure the photons after microwave conversion. The maturity of this technology enables precise measurement of microwaves with a sensitivity on the order of tens of microwave photons, which can be used for detection and imaging of weak microwave photons.

\begin{figure}[htbp]
\centering
\includegraphics[width=1\linewidth]{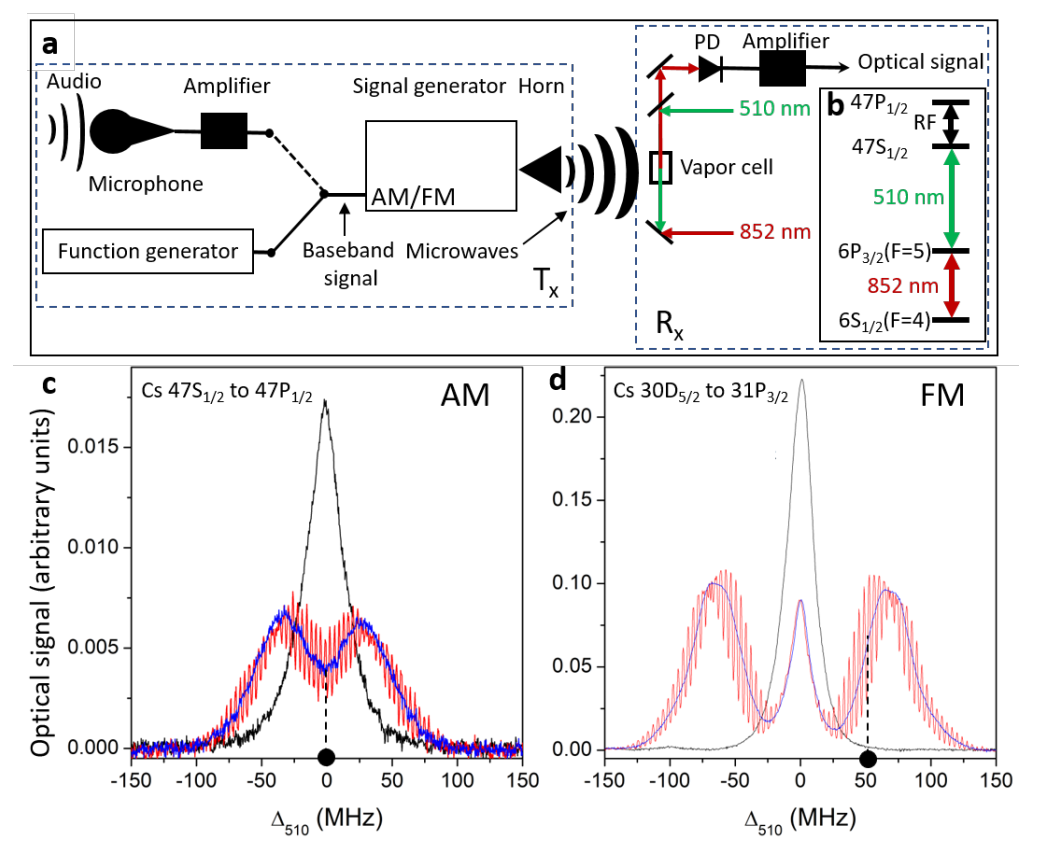}
\caption{(a) Experimental setup. (b) Energy level diagram. (c) EIT spectrum under amplitude modulation (AM) (red), with the carrier amplitude modulated at a baseband AM frequency of 1 kHz and modulation depth of ±25 $\%$. (d) EIT spectrum under frequency modulation (FM) (red), with the carrier frequency modulated at a baseband FM frequency of 1 kHz and modulation deviation of ±30 MHz \cite{39}.}
\label{fig3.1}
\end{figure}

\subsection{Terahertz Electric Field Measurement}
The terahertz band is suitable for nondestructive testing because of its ability to penetrate materials like papers, plastics, and clothes\cite{Jansen:10,https://doi.org/10.1002/lpor.201000011}. Detectors in the THz region are usually less sensitive than those in other regions of the EM spectrum. Researchers at Durham University have demonstrated a THz imaging system based on THz-to-optical conversion in a room-temperature atomic vapor \cite{PhysRevX.10.011027}. This system allows us to acquire images at high speed and sensitivity. Furthermore, many adjustments can still be made in the future to improve its performance.

Electric dipole transitions between neighboring Rydberg states of alkali atoms lie in the THz range\cite{10,58}. THz-to-optical conversion is achieved using the fact that laser-excited Cs Rydberg states ($13D_{5/2}$) can emit a photon at 535 nm when the Rydberg atoms absorb a THz photon, in which the conversion efficiency is approximately 52.4\%. Then, when a 2D sheet of atoms is created, these emitted photons can be used to display a full-field image of the incident THz field in a single exposure. In this experiment \cite{PhysRevX.10.011027}, the system was used to capture the free fall of a water droplet, and the THz imaging of the dynamic process in a frame rose to 3\,kHz. This work shows that the system achieves resolution near the diffraction limit and is capable of high-speed imaging.

\section{Application}
\label{sec4}

A Rydberg atom is very promising in its application owing to its advantages in electric field measurement \cite{83,84,85,86}, especially its extremely wide detection frequency range and high sensitivity. Researchers have put great effort into expanding its applications \cite{80}, as discussed in the following.

\begin{figure*}
\centering
    \includegraphics[width=0.9\linewidth]{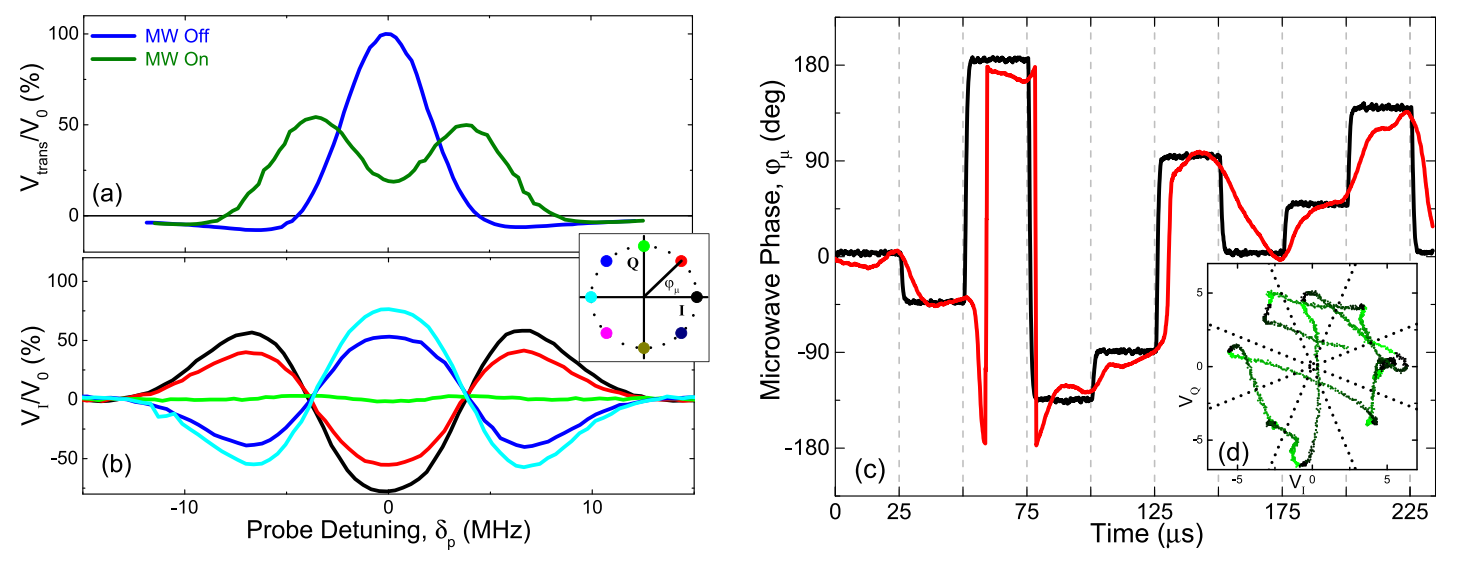}
\caption{(a) Rydberg EIT (blue) and Rydberg EIT-AT splitting (green). (b) Examples of demodulated transmission signals with each color corresponding to a different AM phase. (c) PSK sent and received phases (black and red, respectively). (d) Phase constellation of the received phase in (c) (red line); more details are in \cite{42}.}
\label{fig3.2}
\end{figure*}

\subsection{Communication}

Microwave electric fields are widely used in communication because of their large channel capacity, fast transmission rate, and long transmission distance. 5G communication is an example of application in the microwave band. Rydberg atoms can be used to receive signals because they respond to microwaves \cite{63,64}. 

\subsubsection{Analog Communication}

Atom-based microwave communication is accessible through the microwave electric field measurement introduced above \cite{79}. To realize microwave communication, the baseband information needs to be modulated to the microwave, and then the Rydberg atom serves as a microwave receiver and demodulates the baseband signal.

The carrier electric field $E_c$ is written as
\begin{equation}
    E_c=E_0\rm{cos}(\omega_0t). \label{10}
\end{equation}
The modulated electric field $E_s$ is
\begin{equation}
    E_s=E_1\rm{cos}(\omega_1t). \label{11}
\end{equation}
Then the modulated carrier electric field can be expressed as
\begin{equation}
    E(t)=E_0(1+\frac{E_1}{E_0}\rm{cos}(\omega_1t))\cdot \rm{cos}(\omega_0t), \label{12}
\end{equation}
where $\omega_0$ is the carrier frequency, and $\omega_1$ is the modulation frequency.
The modulated carrier amplitude can be written as
\begin{equation}
    \lvert E(t) \rvert =E_0+E_1\rm{cos}(\omega_1t). \label{13}
\end{equation}

In the EIT readout scheme, the transmittance of the probe light is
\begin{equation}
    T_{probe}\propto \lvert E(t) \rvert =E_0+E_1\rm{cos}(\omega_1t). \label{14}
\end{equation}

We can see that when the EIT scheme is used to read out, the Rydberg atom has the ability to remove the carrier wave. The high-frequency carrier resonates with the Rydberg atomic energy level, and its frequency components are eliminated. The result is that only the low-frequency modulation signal remains in the spectral signal. Therefore, one advantage of using Rydberg atoms for microwave communication is that there is no need for a complex demodulation device. The Rydberg atom can complete the demodulation itself. It filters out the carrier wave and directly demodulates the baseband signal in the probe light signal. To restore the baseband signal, one needs to measure the intensity change of the probe light.

Using the above principles, Georg Raithel's group experimentally demonstrated AM and frequency modulation (FM) communication using Rydberg atoms \cite{39,41}, which can receive, replay, and record baseband signals in the audio range. They used the audio signal to modulate the amplitude or frequency of the microwave carrier wave. The modulated microwaves are detected by the atoms, and the microwaves are demodulated through direct real-time detection of the atomic EIT spectra. Experiments have shown that multi-band AM and FM communication is possible using the rich energy levels of Rydberg atoms. The atomic radio receiver eliminates the need for conventional demodulation and signal-conditioning circuits, making the atomic radio inherently resistant to electromagnetic interference. In addition, the 3-dB bandwidth for communication using this atomic radio receiver is approximately 100 kHz, and is limited by the EIT linewidth, transit time, and other factors \cite{54}. The instantaneous bandwidth of the atomic receiver is only a few hundred kHz, which is shorter than that in conventional communications. This is certainly more than sufficient for transmitting sound signals, but not for efficiently transmitting information at higher frequencies.

\begin{figure*}[htbp]
\centering
\includegraphics[width=0.9\linewidth]{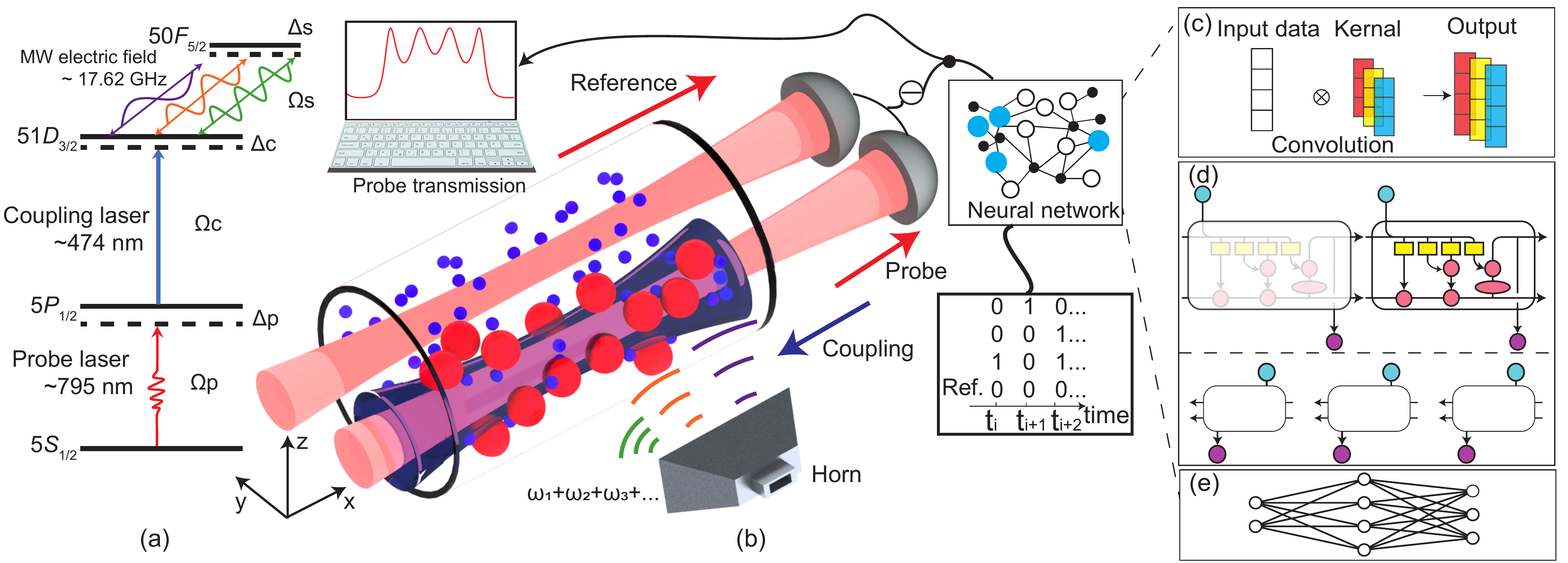}
\caption{\textbf{Illustration of the setup.} (a) Experimental energy diagram. (b) Schematic of a Rydberg atom-based antenna and mixer interacting with multi-frequency signals. The probe light and the counter-propagating coupling light form the EIT configuration. Multi-frequency microwave fields transmitted by a horn are applied to the atoms, with a radiating direction that is perpendicular to the laser beam propagation direction. The multi-frequency microwave fields are modulated using a phase signal carrying the messages. The probe transmission spectrum is fed into a well-trained neural network to retrieve the variations in the phases with time. (c–e) Schematics of the neural network. The network consists of (c) a one-dimensional convolution layer, (d) a bi-directional long-short term memory layer, and (e) a dense layer \cite{Liu2022Deep}.}
\label{Deep}
\end{figure*}

C. L. Holloway's group used Rydberg atoms in real-time recording of musical instruments \cite{40}. In addition, they recorded two guitars simultaneously using two different atomic elements (Cs and Rb) in the same vapor cell. Each atomic element can be detected and can record each guitar separately.

\subsubsection{Digital Communications}

Rydberg atoms can also be used to measure the phase of the microwave electric field through the phase modulation of the field, and then the Rydberg atoms can be used to measure the microwave electric field to achieve digital communication \cite{67,87}. Paul's group demonstrated that room temperature Rydberg atoms can be used as sensitive, high-bandwidth microwave communication antennas \cite{42}. Using an EIT detection scheme, they read out data encoded in AM 17-GHz microwaves near the photon shot noise limit. The channel capacity was up to 8.2 Mbit/s, and an 8-state phase-shift keying (PSK) digital communication protocol was implemented.

For digital microwave communication based on the Rydberg atomic system, the group also studied the standard quantum limit of theoretical data capacity, and experimentally observed quantum-limited data reception with bandwidths from 10 kHz to 30 MHz. This work provides an alternative to microwave communication, which relies on conventional antennas. The efficiency of traditional communication antennas is significantly reduced especially when the antenna is significantly smaller than the electromagnetic field wavelength. However, the efficiency of the Rydberg atomic system is not affected by antenna size \cite{43}.

A group has also studied the feasibility of digital communication with non-resonant continuously tunable RF carriers through Rydberg atomic receivers \cite{44}. Their experiment shows that digital communication at a rate of 500 kbps can be reliably achieved within a tunable bandwidth of 200 MHz near a 10.22-GHz carrier wave. This experiment solidifies the physical foundation for reliable communication and spectral sensing with wider-broadband RF carriers, paving the way for concurrent multi-channel communication based on the same pair of Rydberg states.

\subsubsection{Deep-Learning-Enhanced Rydberg Atomic Multi-Frequency Microwave Recognition and Communication}

Practically, it is challenging to identify multi-frequency microwave electric fields owing to the complex interference between such fields. The excellent properties of Rydberg atoms make them broadly applicable to multi-frequency measurements in microwave radar and microwave communications. However, Rydberg atoms are not only sensitive to microwave signals, but also to noise from atomic collisions and the environment, which means that the solution of the Lindblad master equation for light–atom interactions is complicated by noise and higher-order terms. The USTC team solved these problems by combining Rydberg atoms with a deep-learning model, demonstrating that deep learning leverages the sensitivity of Rydberg atoms and reduces the effects of noise without solving the master equation \cite{Liu2022Deep}. This receiver directly decodes frequency division multiplexed (FDM) signals without multiple bandpass filters and other complex circuits, as shown in Fig.~\ref{Deep}. In this proof-of-principle experiment, an FDM phase-shift keying signal carrying a noisy QR code was effectively decoded with an information transmission rate of 6 kbps for four bins and an accuracy of 99.32\%. This new technology is expected to benefit Rydberg-based microwave field sensing and communication.

\subsection{Sensing}

The advantages of using Rydberg atoms to measure electric fields extend to electric field sensing \cite{65,78,81}. Other field parameters such as phase and polarization can also be measured using Rydberg atoms. This will greatly expand application of Rydberg atom sensing and pave the way for Rydberg atomic electric field sensing.

\subsubsection{Measurement of Microwave Electric Field Phase}

Rydberg atoms are useful in absolute measurements of RF fields using EIT. However, using Rydberg atoms to measure the phase of the RF field is still very challenging. Measuring the phase of RF fields is an essential part of many applications including antenna metrology, communications, and radar. Holloway's group demonstrated a scheme for measuring the phase of an RF field \cite{38,82,88} using Rydberg atoms as a mixer to down-convert an RF field of 20 GHz to an intermediate frequency of the order of kHz. The phase of the intermediate frequency corresponds directly to the RF field phase. They used this method to measure the phase shift of electromagnetic waves in horn antennas because the antennas are at different distances from the Rydberg atomic sensor. Atom-based RF phase measurements allow measurement of the propagation constant of RF waves to within 0.1\% of theoretical values.
Because the phase of an RF field can be measured, Rydberg atom sensors can also be used to measure the angle of arrival (AOA) \cite{robinson2021determining}. When the Rydberg atomic sensor is used to measure the phase at two different locations in the cell, the AOA of the signal can be obtained with a precision of 2.1$^{\circ}$ using the phase difference.

\subsubsection{Rydberg Atomic Microwave Frequency Comb}

Although high sensitivity can be achieved in atomic superheterodyne measurement of a microwave electric field, the instantaneous bandwidth is relatively narrow, which is a critical factor limiting application of such a receiver. New solutions need to be proposed to extend the real-time measurement range and realize measurement of a signal with a certain spectral width. To solve related problems, the USTC team proposed using a microwave frequency comb (MFC) to measure a microwave electric field \cite{PhysRevApplied.18.014033}. A microwave can be measured precisely using a spectrometer with a Rydberg microwave transition frequency comb modified with multiple microwave fields, in which the modified Rydberg atoms display a comb-like RF transition. The Rydberg MFC spectrum provides instantaneous absolute frequency measurement over a range of 125 MHz and gives the relative phase of single-frequency microwave signals. The method is still valid for microwave signals with a certain spectral width. This experiment facilitates real-time detection of unknown microwave signals in a large range using Rydberg atoms.

\begin{figure}[htbp]
\centering
    \includegraphics[width=0.7\linewidth]{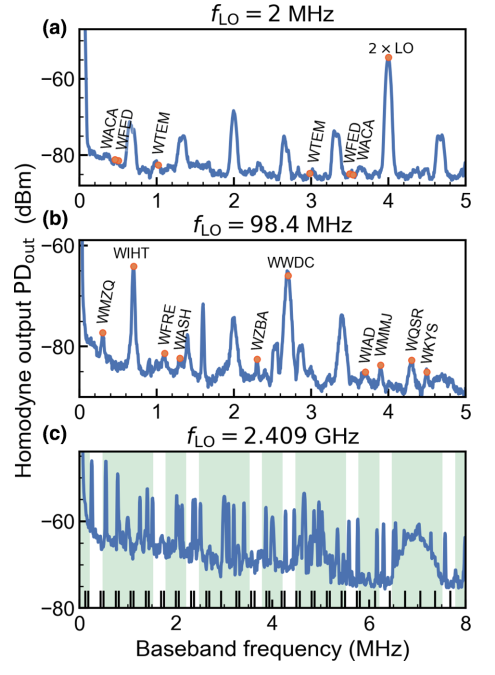}
\caption{RF signals observed in the laboratory using a rabbit ear antenna. AM radio, FM radio, WLAN, and Bluetooth signals can be observed by tuning the LO to $f_{LO}$ = 2 MHz, 98.4 MHz, and 2.409 GHz \cite{45}.}
\label{fig3.3}
\end{figure}

\subsubsection{Large-Bandwidth Measurement of Microwave Electric Fields}

Rydberg atoms respond well to a wide range of microwave electric fields, but continuous measurement is still very challenging. To realize continuous large-bandwidth measurement, Paul's group \cite{45} recently demonstrated an atomic RF receiver and spectrum analyzer based on thermal Rydberg atoms coupled with planar microwave waveguides. Non-resonant RF heterodyne technology was used to achieve continuous operation from DC to the 20 GHz carrier frequency range. The system is DC-coupled and has inherent sensitivity up to $-120(2)$ dBm/Hz, a 4-MHz instantaneous bandwidth, and a linear dynamic range of over 80 dB.

Using a low-noise preamplifier, high-performance spectral analysis has been demonstrated with peak sensitivity better than $-145$ dBm/Hz. Connecting a standard rabbit ear antenna enables the spectrum analyzer to detect weak ambient signals including FM radio, AM radio, WiFi, and Bluetooth.
This waveguide coupling greatly strengthens the coupling between free-space microwaves and Rydberg atoms by restricting the electric field to the region of the Rydberg atoms. The system makes it possible to develop small, room temperature, integration-based Rydberg sensors that exceed the bandwidth. Such sensors may further exceed the thermal noise limit on the sensitivity of conventional RF sensors, receivers, and analyzers.

\subsubsection{Vector Microwave Measurements}

\begin{figure}[htbp]
\centering
    \includegraphics[width=1\linewidth]{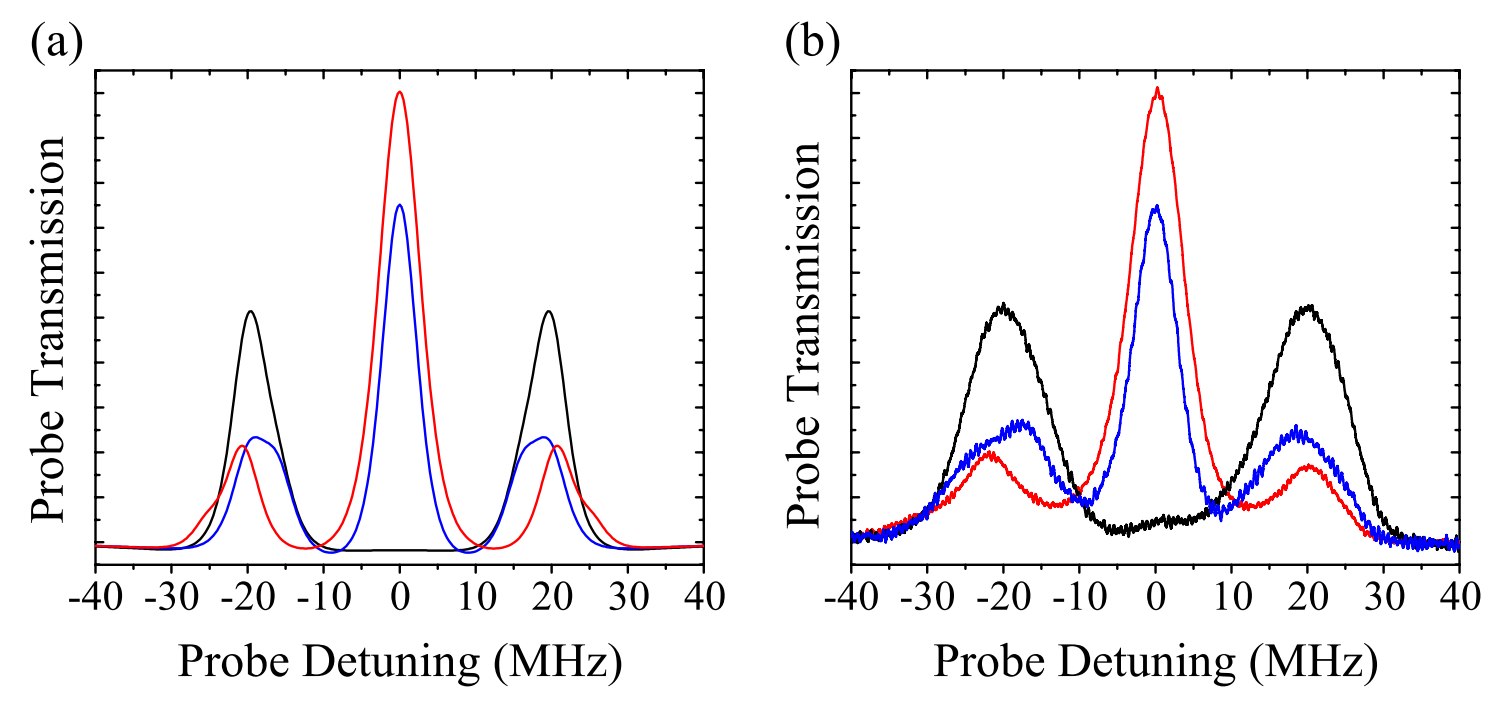}
\caption{EIT spectrum corresponding to microwaves with different polarization directions \cite{47}.}
\label{fig3.4}
\end{figure}

Rydberg atomic EIT-AT spectroscopy can also be used to measure microwave polarization \cite{47,59}. Because the EIT spectrum is sensitive to the laser polarization, the microwave polarization angle relative to the laser affects the shape of the spectrum. When the selected atomic energy level is in a four-level system, different results can be obtained by changing the polarization combination of the probe light, coupling light, and microwave electric field, as shown in Fig.~\ref{fig3.4}. 

The microwave electric field in different directions can be divided into two parts, one of which couples the Rydberg state and the other does not interact with the Rydberg state. The relative strength of these two parts depends on the angle between the microwave electric field polarization and the laser polarization, so changing the microwave polarization direction changes the transmission spectrum of the probe light. The polarization direction of the probe and the coupled fields is given, so the polarization direction of the microwave electric field can be determined, and the angle measured through this method can be precise to $0.5^{\circ}$.

\section{Outlook}
\label{sec4}

Compared with traditional antenna measurement, electric field measurement based on Rydberg atoms has its own unique advantages. Examples are its higher sensitivity, size independence, and large working bandwidth. The interaction between Rydberg atoms and the applied electric field can produce interesting physics and expand limits.

Although there has been much progress in enhancing the sensitivity of Rydberg receivers \cite{89,90}, no reported study so far has exceeded the sensitivity of conventional microwave receivers. The minimum measurable field strength reaches 780 pV/cm \cite{7}, and a sensitivity of 12.5 nV/cm/Hz$^{1/2}$ \cite{cai2022sensitivity} has been reported. The quantum projection noise limit for Rydberg atoms is lower than that of traditional measurement methods, therefore promising higher sensitivity. Further exploration is required \cite{santamaria2022comparison}. In the future, Rydberg atoms can be used to obtain higher sensitivity and then replace traditional metal antennas in microwave applications. 

Continuous working of Rydberg atomic RF sensing is a fundamental requirement to avoid multiple down-conversion for traditional receivers. This is conducive to directly receiving television, satellite, WIFI, mobile phone, and other signals. This is fundamentally different from traditional microwave receiving systems. Electric fields have been detected in several separated frequency bands from the kHz to the THz regions. However, there is much room for developing a larger working bandwidth than that of the traditional receiver. 

Instantaneous bandwidth is also an important parameter for building a radar based on Rydberg atoms. This parameter determines the resolution of object identification or the communication rate. However, it is limited by the relaxation time at which the atomic system reaches a steady state, which always in the range of kHz $\sim$ MHz. This is far from that of the traditional measurement device, which often provides an instantaneous bandwidth on the order of GHz. New physics and technology need to be developed to improve this parameter.

In a traditional antenna measurement system, it is necessary to use different antenna sizes to receive different electromagnetic wavelengths. However, Rydberg atomic receivers can be used for continuous broadband signal measurement without changing the sensing element, which is conducive to device miniaturization and integration. This would help load Rydberg sensors on a satellite for sensing MHz electric fields because traditional MHz radars are big. Thus, this is very promising for sensing microwave signals in a small aircraft.

Future work can demonstrate the advantages of Rydberg atom sensors with multiple indexes at the same time, such as with indexes for both working bandwidth and sensitivity. The advantages of the Rydberg atom may be specific to special areas such as detecting the field distribution near a signal source, in which the particular requirements of miniaturization and non-calibration could be met perfectly by the Rydberg atoms.

\begin{acknowledgments}
We acknowledge funding from the National Key R\&D Program of China (Grant No. 2022YFA140400), the National Natural Science Foundation of China (Grant Nos. U20A20218, 61525504, and 61435011), and the Youth Innovation Promotion Association of the Chinese Academy of Sciences (Grant No. 2018490).
\end{acknowledgments}

\bibliography{refs}

\end{document}